\documentclass[draftcls,onecolumn]{IEEEtran}

\usepackage{amsmath}
\usepackage{amssymb}
\usepackage{latexsym}
\usepackage{multirow}
\usepackage{epsfig}
\usepackage{graphics}
\usepackage{mathrsfs}
\usepackage{algorithmic}
\usepackage{algorithm}

\newcommand{\eqdef}{\stackrel{\triangle}{=}}
\newcommand{\beq}{\begin{equation}}
\newcommand{\enq}{\end{equation}}
\newcommand{\ben}{\begin{eqnarray}}
\newcommand{\enn}{\end{eqnarray}}
\newcommand{\bei}{\begin{itemize}}
\newcommand{\eni}{\end{itemize}}

\newtheorem{theorem}{Theorem}[section]

\begin{document}

\title{Compute-and-Forward Network Coding Design over Multi-Source Multi-Relay Channels}

\author{Lili Wei and WenChen

\thanks{Manuscript received October 31, 2011; revised April 6, 2012; accepted May 18, 2012. The associate editor coordinating the review of this paper and approving it for publication was J. Luo.}
\thanks{The authors are with the Department of Electronic Engineering,
Shanghai Jiao Tong University, China (e-mail: \{liliwei, wenchen\}@sjtu.edu.cn). W. Chen is also with SKL for ISN, Xidian University, China.}
\thanks{This work is supported by the National 973 Project \#2012CB316106,
by NSF China \#60972031 and \#61161130529, by the National 973
Project \#2009CB824904, and by the National Key Laboratory Project
\#ISN11-01.}
\thanks{Digital Object Identifier XXXXXXXXXX}}

\markboth{IEEE Transactions ~2012} {Shell \MakeLowercase{\textit{WEI and CHEN}}:
Compute-and-Forward Network Coding Design over Multi-Source Multi-Relay Channels}
\maketitle

\begin{abstract}

Network coding is a new and promising paradigm for modern communication networks by allowing intermediate nodes to mix messages received from multiple sources.  Compute-and-forward strategy is one category of network coding in which a relay will decode and forward a linear combination of source messages according to the observed channel coefficients, based on the algebraic structure of lattice codes. The destination will recover all transmitted messages if enough linear equations are received. In this work, we design in a system level, the compute-and-forward network coding coefficients by Fincke-Pohst based candidate set searching algorithm and network coding system matrix constructing algorithm, such that by those proposed algorithms, the transmission rate of the multi-source multi-relay system is maximized. Numerical results demonstrate the effectiveness of our proposed algorithms.

\end{abstract}


\IEEEpeerreviewmaketitle

\section{Introduction}

\IEEEPARstart{S}{ince} the pioneering research work of Ahlswede {\em et al.} in 2000 \cite{Cai_NC}, network coding (NC) has rapidly emerged as a major research area in electrical engineering and computer science. NC is a generalized routing approach that breaks the traditional assumption of simply forwarding data, and allows intermediate nodes to send out functions of their received packets, by which the multicast capacity can be achieved. Subsequent works of \cite{Cai_LNC}-\cite{Chou_Poly} made the important observation that, for multicasting, intermediate nodes can simply send out a linear combination of their received packets. Linear network coding with random coefficients is considered in \cite{Medard_Random NC}. Physical layer network coding is presented in \cite{Zhang_PLNC}. Complex field network coding is proposed in \cite{CFNC}. Several other network coding realizations in wireless networks are discussed in \cite{Katti_ANC}-\cite{Medard-wireless}.

There is also a large body of works on lattice codes \cite{Lattice1}-\cite{Lattice2} and their applications in communications. For many AWGN networks of interest, lattice codes with linear structure can approach the performance of standard random coding arguments. It has been shown that nested lattice codes (combined with lattice decoding) can achieve the capacity of the point-to-point AWGN channel \cite{Nested1}-\cite{Nested2}. Also, another appealing aspect of linear lattice codes lies in their lower decoding complexity by a class of efficient decoders \cite{SD0}-\cite{SD3}. In the two-way relay networks, a nested lattice based strategy has been developed that the achievable rate is near the optimal upper bound \cite{Lattice-TWRC1}-\cite{Lattice-TWRC3}.

Recently, a new strategy of compute-and-forward (CPF) \cite{CPF}-\cite{Reliable PLNC}, beneficial from both network coding and lattice codes, attracts great attention. The main idea is that a relay will decode a linear function of transmitted messages according to the observed channel coefficients rather than ignoring the interference as noise. Upon utilizing the algebraic structure of lattice codes, i.e. the integer combination of lattice codewords is still a codeword as well, the intermediate relay node decodes and forwards an integer combination of original messages. With enough linear independent equations, the destination can recover the original messages respectively. Subsequent works for design and analysis of the CPF strategy have been given in \cite{CPF1}-\cite{CPF4}. The idea of MIMO compute-and-forward is presented in \cite{MIMO CPF}.

Those previous works in CPF only consider the integer network coding coefficients optimization of each relay locally/separately. However, for a multi-source multi-relay system with $L$ sources, the previous separate optimizations cannot guarantee the network coding system matrix, which is constructed by all the integer network coding coefficient vectors, is of rank $L$ such that the destination can decode all messages. In this work, the compute-and-forward network coding strategy is considered in a system level. First, by our proposed Fincke-Pohst \cite{SD0} based candidate set searching algorithm, instead of one optimal network coding coefficient vector, for each relay we will provide a network coding vector candidate set with corresponding computation rate in descending order. Then, by our proposed network coding system matrix constructing algorithm, we will try to choose network coding vectors from those candidate sets to construct network coding system matrix with rank $L$, while in the meantime the transmission rate of the multi-source multi-relay system is maximized. The underlying codes are based on lattice codes whose algebraic structure ensures that integer combinations of messages can be decoded reliably.

The notations used in this work are as follows. $\{\cdot\}^T$ denotes the transpose operation, $|\cdot|$ represents the cardinality of a set, $\mathbb Z^n$ denotes the $n$ dimensional integer ring, $\mathbb R^n$ denotes the $n$ dimensional real field, $\mathbb F_p$ denotes a finite field of size p. $\mathbf I_n$ denotes the identity matrix of size $n\times n$, and $\mathbf 0$ denotes the vectors with all zeros elements. Assume that the $\log$ operation is with respect to base $2$. We use boldface lowercase letters to denote column vectors and boldface uppercase letters to denote matrices.

\section{Multi-Source Multi-Relay Channel}

\subsection{System Model}

We consider the multi-source multi-relay (MSMR) system model as shown in Fig. 1, where $L$ sources $\mathcal{S}_1$, $\mathcal{S}_2$, $\cdots$, $\mathcal{S}_L$ are communicating to one destination $\mathcal{D}$ through $L$ relays $\mathcal{R}_1$, $\mathcal{R}_2$, $\cdots$, $\mathcal{R}_L$. Each node is equipped with a single antenna and works in half-duplex mode. There are no direct links from sources to the destination.

\begin{figure}[hbt]
\centerline{\psfig{file=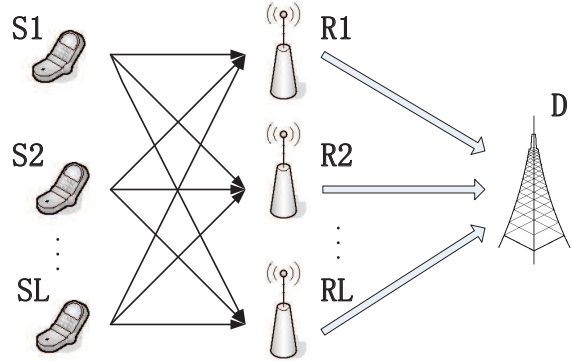,width=3in}}
\centering
\caption{System model of a MSMR network}
\label{figure-MSMR_system}
\end{figure}

The information transmission, which we call one transmission realization, is performed in two phases. The first phase is for the transmissions from all sources $\mathcal{S}_1$, $\mathcal{S}_2$, $\cdots$, $\mathcal{S}_L$ to the relays $\mathcal{R}_1$, $\mathcal{R}_2$, $\cdots$, $\mathcal{R}_L$. Each relay will receive signals from all sources due to the wireless medium. In the second phase, assume each relay has a point-to-point AWGN channel or orthogonal access to the destination, for example, in different time slots as shown in Fig. 2. Every relay will obtain a linear combination of original messages and forward towards the destination by orthogonal channels. With enough linear combinations, the destination is able to recover the desired original messages from all sources.

\begin{figure}[hbt]
\centerline{\psfig{file=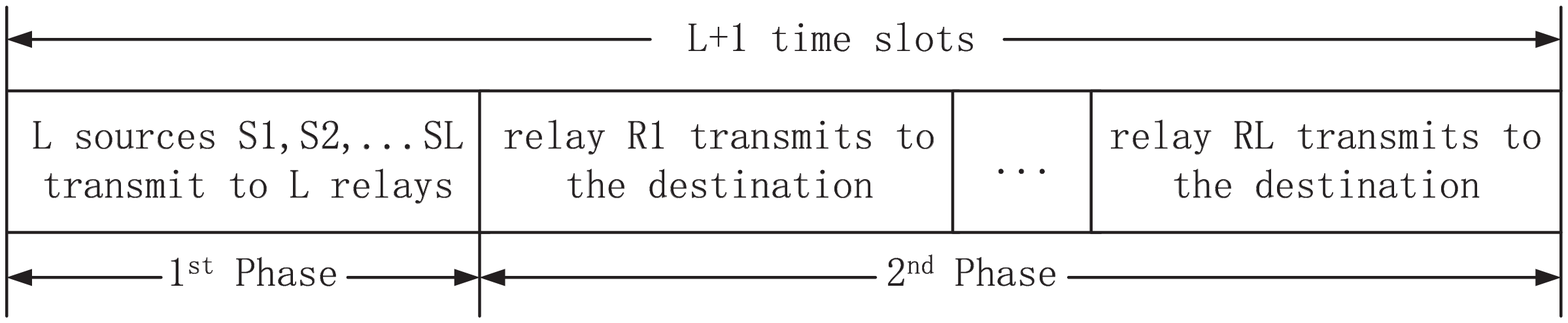,width=3.5in}}
\centering
\caption{Time division allocation for one transmission realization}
\label{figure-time-schedule}
\end{figure}

Without loss of generality, in one transmission realization, each source has a length-$k$ message vector that is drawn independently and uniformly over a prime size finite field,
\beq
\mathbf w_l\in\mathbb F_p^k,\quad l=1,2,\cdots,L,
\enq
where $\mathbb F_p$ denotes the finite field with a set of $p$ elements. Each source is equipped with an encoder $\Psi_l: \mathbb F_p^k \to {\mathbb R}^n$ that maps the length-$k$ message $\mathbf w_l$ into a length-$n$ real valued lattice codeword ${\bf x}_l=\Psi_l(\mathbf w_l)$. The lattice codeword $\mathbf x_l$ must satisfy the power constraint, $\frac{1}{n}||{\bf x}_l||^2\leq P$ for $P\ge 0$ and $l=1,2,\cdots,L$. The message rate, defined as the length of the message measured in bits normalized by the number of channel uses $R=\frac{k}{n}\log p$ \cite{CPF}, is the same for all sources.

After mapping its message $\mathbf w_l\in\mathbb F_p^k$ into a lattice codeword ${\bf x}_l\in \mathbb R^n$, the source $\mathcal{S}_l$ will send the codeword $\mathbf x_l$ across the channel. Due to the broadcast nature of wireless medium, the $m$-th relay will observe a noisy combination of the transmitted signals at the end of the first phase,
\beq
{\bf y}_{m} = \sum_{l=1}^{L} h_{ml}\mathbf x_l + \mathbf z_m\label{ym},\quad\quad m=1,2,\cdots,L,
\enq
where $h_{ml}\in \mathbb R$ denotes real valued fading channel coefficient from $\mathcal{S}_l$ to relay $\mathcal{R}_m$, generated i.i.d. according to a normal distribution $\mathcal N (0,1)$; ${\bf z}_{m}\in \mathbb R^{n}$ denotes additive Gaussian noise vector, $\mathbf z_m \sim \mathcal{N}(\mathbf 0,\mathbf I_{n})$. Let
\beq
\mathbf h_m = [h_{m1}, \cdots, h_{mL}]^T
\enq
denote the vector of channel coefficients from all sources to the $m$-th relay. We assume this channel state information $\mathbf h_m$ is available at relay $m$.

\subsection{Compute-and-Forward Scheme}

In a recent work, Nazer and Gastpar propose the {\em compute-and-forward} approach \cite{CPF} which exploits the property that any integer combination of lattice points is again a lattice point. After receiving the noisy vector $\mathbf y_m$ of (\ref{ym}), the $m$-th relay will first select a scalar $\beta_m\in\mathbb R$ and an integer network coding coefficient vector $\mathbf a_m = [a_{m1},a_{m2},\cdots,a_{mL}]^T\in\mathbb Z^L$, then attempt to decode the lattice point $\sum_{l=1}^L a_{ml}\mathbf x_l$ from
\ben
\beta_m \mathbf y_m & = & \sum_{l=1}^L \beta_m h_{ml} \mathbf x_l + \beta_m\mathbf z_m\\
& = & \sum_{l=1}^L a_{ml}\mathbf x_l + \underbrace{\sum_{l=1}^L \left(\beta_m h_{ml} - a_{ml}\right)\mathbf x_l + \beta_m\mathbf z_m}_{\textit{Effective Noise}}.\quad\quad
\enn
Note that we do not need to conduct joint maximum likelihood (ML) decoding to get $(\hat{\mathbf x}_1,\hat{\mathbf x}_2,\cdots,\hat{\mathbf x}_L)$ for network coding. Instead we decode $\sum_{l=1}^L a_{ml}\mathbf x_l$ as one regular codeword due to the lattice algebraic structure. In other words, the network coded codeword is still in the same field as original source codeword.

In the finite field, it is equivalent that each relay is desired to reliably recover a linear combination of the messages,
\beq
\mathbf u_m = \bigoplus_{l=1}^L q_{ml}\mathbf w_l = \left[\sum_{l=1}^L a_{ml}\mathbf w_l\right] mod\; p,
\enq
where $\bigoplus$ denotes summation over the finite field, $q_{ml}$ is a coefficient taking values in $\mathbb F_p$ and $q_{ml} = a_{ml}\; mod \;p$.

Each relay is equipped with a decoder, $\Pi_m: \mathbb R^n \to \mathbb F_p^k$, that maps the observed channel output $\mathbf y_m\in \mathbb R^n$ to an estimate $\hat{\mathbf u}_m =\Pi_m(\mathbf y_m) \in \mathbb F_p^k$ of the message combination $\mathbf u_m$. The diagram of compute-and-forward scheme is given in Fig. 3.

\begin{figure}[hbt]
\centerline{\psfig{file=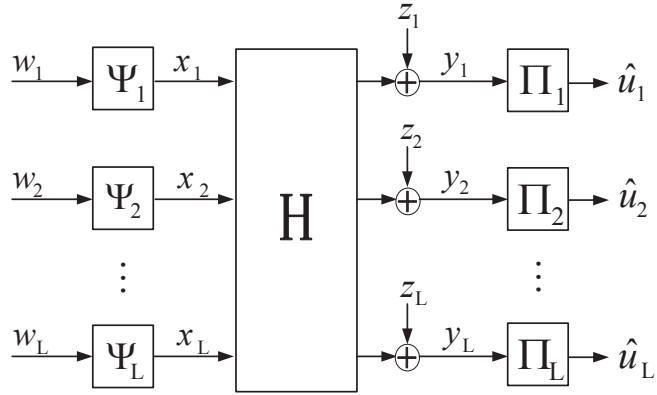,width=3.5in}}
\centering
\caption{Compute-and-Forward Diagram}
\label{figure-CPF-diagram}
\end{figure}

We are interested in the rate of $\sum_{l=1}^L a_{ml}\mathbf x_l$ as a whole and will capture the performance of the computation scheme by what we refer to as the {\em computation rate}, namely, the number of bits of the linear function successfully recovered per channel use. The work of \cite{CPF} shows that a relay can often recover an equation of messages at a higher rate than any individual message (or subset of message). The rate is highest when the equation coefficients closely approximate the effective channel coefficients. The formal statements are given in the following theorems \cite{CPF}-\cite{CPF1}. Let $\log^+(x) \eqdef \max(\log(x),0)$.\\

\begin{theorem}\label{theorem-computation rate}
For real-valued AWGN networks with channel coefficient vector $\mathbf h_m\in\mathbb R^L$ and desired network coding coefficient vector $\mathbf a_m\in\mathbb Z^L$, the following computation rate is achievable
\beq
\mathscr R_m(\mathbf a_m) = \max_{\beta_m\in \mathbb R} \frac{1}{2}\log^+\left(\frac{P}{\beta_m^2 + P ||\beta_m \mathbf h_m -\mathbf a_m||^2}\right).
\enq
\end{theorem}

\begin{theorem}\label{theorem-alpha}
The computation rate given in Theorem \ref{theorem-computation rate} is uniquely maximized by choosing $\beta_m$ to be the MMSE coefficient
\beq
\beta_{MMSE} = \frac{P\,\mathbf h_m^T \mathbf a_m}{1+P||\mathbf h_m||^2},
\enq
which results in a computation rate of
\beq
\mathscr R_m(\mathbf a_m) = \frac{1}{2}\log^+\left(||\mathbf a_m||^2 - \frac{P (\mathbf h_m^T \mathbf a_m)^2}{1 + P ||\mathbf h_m||^2}\right)^{-1}.\label{Rm}
\enq
\end{theorem}

\begin{theorem}\label{theorem-am}
For a given channel coefficient vector $\mathbf h_m = [h_{m1},h_{m2},\cdots,h_{mL}]^T\in\mathbb R^L$, $\mathscr R_m(\mathbf a_m)$ is maximized by choosing the integer network coding coefficient vector $\;\mathbf a_m\in\mathbb Z^L$ as
\beq
\mathbf a_m = \textit{arg}\min_{\mathbf a_m\in\mathbb Z^L,\mathbf a_m\neq \mathbf 0}\left(\mathbf a_m^T \mathbf G_m \mathbf a_m\right),\label{am0}
\enq
where
\beq
\mathbf G_m \eqdef \mathbf I - \frac{P}{1 + P\,||\mathbf h_m||^2} \mathbf H_m,\label{Gm}
\enq
and $\mathbf H_m = [H^{(m)}_{ij}]$, $H^{(m)}_{ij}=h_{mi} h_{mj}$, $1\le i,j\le L$.
\end{theorem}

\subsection{Problem Statement}

Theorems \ref{theorem-computation rate}-\ref{theorem-am} only give the optimal network coding integer coefficient vector $\mathbf a_m$ and achievable computation rate $\mathscr R_{m}$ for each relay locally/separately and do not take consideration of the overall system constraints. For the multi-source multi-relay system, at the destination, enough linear combinations of the original messages need to be collected. Let $\mathbf a_1$, $\mathbf a_2$, $\cdots$, $\mathbf a_L$ be the integer network coding coefficients vector for each relay, then the network coding system matrix $\mathbf A$ at the destination can be denoted as
\beq
\mathbf A = \left[\mathbf a_1, \mathbf a_2, \cdots, \mathbf a_L\right]^T.
\enq
Hence, the destination can solve for the original packets if the network coding system matrix $\mathbf A$ has full rank $L$, i.e. $|\mathbf A|\neq 0$. In which case, as the same rate of source-relay channels in phase I is available for relay-destination channels in phase II, the transmission rate at the destination is dominated/bottlenecked by
\beq
\mathscr R_{D} = \min\left\{\mathscr R_{1},\mathscr R_2,\cdots, \mathscr R_L\right\}.
\enq

We can easily understand that after calculating the integer network coding coefficient vector $\mathbf a_m$ for each relay by theorems \ref{theorem-computation rate}-\ref{theorem-am} to maximize its own computation rate, the network coding system matrix $\mathbf A$ constructed by those integer vectors may not have full rank $L$, in which case the destination cannot decode the original messages by those linear equations. In other words, we cannot fix the optimal integer network coding vector $\mathbf a_m$ for each relay separately, since it cannot guarantee that the system constraint of full rank $\mathbf A$.

Therefore, we need to optimize the integer network coding vectors for $L$ relays in a overall system level. Instead of distributed calculations, to construct the full rank network coding system matrix that maximize the overall message rate at destination, $\mathbf A$ will be designed according to the following criteria\\
\makebox
{\scalebox{0.92}{%
\parbox{0.55\textwidth}{\ben
\mathbf A & = & \textit{arg}\,\max_{|\mathbf A|\neq 0}{\mathscr R_{D}}\nonumber\\
 & = & \textit{arg}\,\max_{|\mathbf A|\neq 0}\left( \min\left\{\mathscr R_{1},\mathscr R_2,\cdots, \mathscr R_L\right\}\right)\nonumber\\
 & = & \textit{arg}\,\max_{|\mathbf A|\neq 0}\;\min_{m=1,\cdots L}\; \left(\frac{1}{2}\log^+\left(||\mathbf a_m||^2 - \frac{P (\mathbf h_m^T \mathbf a_m)^2}{1 + P ||\mathbf h_m||^2}\right)^{-1}\right).\nonumber\\
 & & \label{RD}
\enn}}}\\
In other words, we need to find the integer network coding vectors $\mathbf a_1$, $\mathbf a_2$, $\cdots$, $\mathbf a_L$, under the system level constraint of full rank $\mathbf A$,  to maximize the computation rate of each relay $\mathscr R_1$, $\mathscr R_2$, $\cdots$, $\mathscr R_L$ jointly, such that the minimum value of $\mathscr R_1$, $\mathscr R_2$, $\cdots$, $\mathscr R_L$ is maximized.

Equivalently, the optimum network coding system matrix $\mathbf A$ should be
\beq
\mathbf A = \textit{arg}\,\min_{|\mathbf A|\neq 0}\;\max_{m=1,\cdots L}\; \mathbf a_m^T \mathbf G_m \mathbf a_m,\label{A}
\enq
where $\mathbf G_m$ is defined in (\ref{Gm}).

\section{Proposed Strategy}

In this work, to approach the overall system optimization of (\ref{RD})-(\ref{A}), we propose the following novel strategy which includes two steps. In the first step, for relay $m$, instead of finding one optimal network coding coefficient vector $\mathbf a_m$ to maximize its own computation rate, we are trying to find a candidate set
\beq
\Omega_m^{T_{max}}=\{\mathbf a_m^{(1)}, \mathbf a_m^{(2)}, \cdots, \mathbf a_m^{(T_{max})}\},
\enq
with $|\Omega_m^{T_{max}}|=T_{max}$. The network coding coefficient vectors with the top $T_{max}$ maximum computation rates for relay $m$ are within the candidate set $\Omega_m^{T_{max}}$. Note that $T_{max}$ is a parameter to control the candidate set length for each relay and currently set by experience/simulation. We will propose an algorithm based on Fincke-Pohst Method \cite{SD0} to find the network coding coefficient vector candidate set for each relay.

After we get all the candidate vector sets $\Omega_1^{T_{max}}$, $\Omega_2^{T_{max}}$, $\cdots$, $\Omega_L^{T_{max}}$, in the second step, we will try to pick up $\mathbf a_1\in \Omega_1^{T_{max}}$, $\mathbf a_2\in \Omega_2^{T_{max}}$, $\cdots$, $\mathbf a_L\in \Omega_L^{T_{max}}$, to construct the full rank network coding coefficient matrix $\mathbf A = \left[\mathbf a_1, \mathbf a_2, \cdots, \mathbf a_L\right]^T$, while in the meantime, the minimum value of corresponding $\mathscr R_1(\mathbf a_1)$, $\mathscr R_2(\mathbf a_2)$, $\cdots$, $\mathscr R_L(\mathbf a_L)$ is maximized.

\subsection{Searching Candidate Set $\Omega_m^{T_{max}}$ for Each Relay}

For relay $m$, we are trying to find the candidate set $\Omega_m^{T_{max}}=\{\mathbf a_m^{(1)}, \mathbf a_m^{(2)}, \cdots, \mathbf a_m^{(T_{max})}\}$ with $|\Omega_m^{T_{max}}|=T_{max}$, such that the network coding coefficient vectors with the top $T_{max}$ maximum computation rate for relay $m$ are within. According to Theorem \ref{theorem-am}, it is equivalent to find the set $\Omega_m^{T_{max}}$ with $T_{max}$ vectors, such that those vectors give the bottom $T_{max}$ minimum $\mathbf a_m^T \mathbf G_m \mathbf a_m$ values, where $\mathbf G_m$ is defined in (\ref{Gm}).

The searching of candidate set $\Omega_m^{max}$ with fixed length $T_{max}$ can be decomposed into following steps.
\bei
\item[(1)] Enumerate all vectors $\mathbf t\in \mathbb Z^L$ ($\mathbf t\neq \mathbf 0$) in $\Omega_m$, such that $\mathbf t^T \mathbf G_m \mathbf t\le C$ for a given positive constant C, i.e.,
\beq
\Omega_m = \left\{\mathbf t: \;\mathbf t^T \mathbf G_m \mathbf t\le C, \;\mathbf t\neq \mathbf 0,\;\mathbf t\in \mathbb Z^L\right\}.\label{Omega}
\enq
\item[(2)] Adjust the constant C to guarantee that $|\Omega_m|\ge T_{max}$.
\item[(3)] Sort all the vectors $\mathbf t_1$, $\mathbf t_2$, $\cdots$, $\mathbf t_{|\Omega_m|}$ in $\Omega_m$ in descending order corresponding to the computation rate value $\mathscr R_m$ in (\ref{Rm}), such that
\beq
\mathscr R_m(\mathbf t_1) \ge \mathscr R_m(\mathbf t_2) \ge \cdots \ge \mathscr R_m(\mathbf t_{|\Omega_m|}).
\enq
\item[(4)] Pick the first $T_{max}$ vectors of $\Omega_m$ to form the set $\Omega_m^{T_{max}}$.
\eni


The procedure of enumerating all vectors $\mathbf t\in \mathbb Z^L$ ($\mathbf t\neq \mathbf 0$) in $\Omega_m$, such that $\mathbf t^T \mathbf G_m \mathbf t\le C$ for a given positive constant C is based on the Fincke-Pohst Method and derived as follows.

We operate Cholesky's factorization of matrix ${\bf G}_m$, ${\bf G}_m = {\bf U}^T {\bf U}$, where ${\bf U}$ is an upper triangular matrix. Denote $||\cdot||_F$ for the Frobenius norm. Let $u_{ij}$, $i,j=1,2,\cdots,L$, be the entries of the upper triangular matrix ${\bf U}$ and $\mathbf t=[t_1,t_2,\cdots,t_L]^T$. Then, the searching vector $\mathbf t$ that makes $\mathbf t^T \mathbf G_m \mathbf t\le C$ can be expressed as\\
\makebox
{\scalebox{0.85}{%
\parbox{0.6\textwidth}{\ben
{\bf t}^T {\bf G}_m {\bf t} & = & ||{\bf U}\;{\bf t}||_F^2 = \sum_{i=1}^{L} \left(u_{ii}t_{i} + \sum_{j=i+1}^L u_{ij}t_{j}\right)^2\nonumber\\
& = & \sum_{i=1}^{L} g_{ii} \left(t_{i} + \sum_{j=i+1}^L g_{ij}t_{j}\right)^2\nonumber\\
& = & \sum_{i=k}^{L} g_{ii} \left(t_{i} + \sum_{j=i+1}^L g_{ij}t_{j}\right)^2 + \sum_{i=1}^{k-1} g_{ii} \left(t_{i} + \sum_{j=i+1}^L g_{ij}t_{j}\right)^2\nonumber\\
&\le & C\label{C_all}
\enn}}}\\
where $g_{ii}=u_{ii}^2$ and $g_{ij}=u_{ij}/u_{ii}$ for $i = 1, 2, \cdots,L$, $j = i+1, \cdots, L$. Obviously the second term of (\ref{C_all}) is non-negative, hence, to satisfy (\ref{C_all}), it is equivalent to consider for every $k = L, L-1, \cdots, 1$,
\beq
\sum_{i=k}^{L} g_{ii} \left(t_{i} + \sum_{j=i+1}^L g_{ij}t_{j}\right)^2 \le C. \label{arg_C}
\enq
Then, we can start work backwards to find the bounds for vector entries $t_{L}, t_{L-1}, \cdots, t_{1}$ one by one.

We begin to evaluate the last element $t_L$ of the searching vector ${\bf t}$. Referring to (\ref{arg_C}) and let $k=L$, we have
\beq
g_{LL}t_L^2 \le C.\label{sL0}
\enq
Set $\Delta_L=0$, $C_L=C$, and we will get
\beq
LB_L \le t_L \le UB_L,\label{sL1}
\enq
with
\ben
UB_L = \left\lfloor\; \sqrt{\frac{C_L}{g_{LL}}} - \Delta_L \;\right\rfloor, LB_L = \left\lceil\; -\sqrt{\frac{C_L}{g_{LL}}} - \Delta_L \; \right\rceil,\label{sL2}
\enn
where $\lceil x \rceil$ is the smallest integer no less than $x$ and $\lfloor x \rfloor$ is the greatest integer no bigger than $x$.

Next, we evaluate the element $t_{L-1}$ of the searching vector ${\bf t}$. Referring to (\ref{arg_C}) and let $k=L-1$, we have
\beq
g_{LL} t_{L}^2 + g_{L-1,L-1} \left(t_{L-1} + g_{L-1,L}t_{L}\right)^2 \le C,
\enq
which leads to\\
\makebox
{\scalebox{0.85}{%
\parbox{0.5\textwidth}{
\beq
\left\lceil\; -\sqrt{\frac{C - g_{LL} t_L^2}{g_{L-1,L-1}}} - g_{L-1,L} t_L \;\right\rceil \le t_{L-1} \le \left\lfloor\; \sqrt{\frac{C - g_{LL} t_L^2}{g_{L-1,L-1}}} - g_{L-1,L} t_L \;\right\rfloor.
\enq}}}\\
If we denote $\Delta_{L-1} = g_{L-1,L} t_L$, $C_{L-1} = C - g_{LL} t_L^2$, the bounds for $s_{L-1}$ can be expressed as
\beq
LB_{L-1} \le t_{L-1} \le UB_{L-1},\label{sL-11}
\enq
where\\
\makebox
{\scalebox{0.8}{%
\parbox{0.5\textwidth}{
\ben
UB_{L-1} = \left\lfloor\sqrt{\frac{C_{L-1}}{g_{L-1,L-1}}} - \Delta_{L-1}\right\rfloor, LB_{L-1} = \left\lceil -\sqrt{\frac{C_{L-1}}{g_{L-1,L-1}}} - \Delta_{L-1}\right\rceil.
\label{sL-12}
\enn}}}\\
We can see that given radius $\sqrt{C}$ and matrix ${\bf U}$, the bounds for $t_{L-1}$ only depends on the previous evaluated $t_{L}$, and not correlated with $t_{L-2}, t_{L-3}, \cdots, t_{1}$.

In a similar fashion, we can proceed for $t_{L-2}$ evaluation, and so on.

To evaluate the element $t_k$ of the searching vector ${\bf t}$, referring to (\ref{arg_C}) we will have
\beq
\sum_{i=k}^{L} g_{ii} \left(t_i + \sum_{j=i+1}^L g_{ij}t_j\right)^2 \le C,
\enq
which leads to\\
\makebox
{\scalebox{0.8}{%
\parbox{0.5\textwidth}{
\ben
& \left\lceil\; -\sqrt{\frac{1}{g_{kk}} \left(C - \sum_{i=k+1}^L g_{ii}\left(t_i + \sum_{j=i+1}^L g_{ij}t_j\right)^2\right)} - \sum_{j=k+1}^L g_{kj} t_j \;\right\rceil\nonumber\\
& \le t_k \le \left\lfloor\; \sqrt{\frac{1}{g_{kk}} \left(C - \sum_{i=k+1}^L g_{ii}\left(t_i + \sum_{j=i+1}^L g_{ij}t_j\right)^2\right)} - \sum_{j=k+1}^L g_{kj} t_j \;\right\rfloor.\nonumber
\enn}}}\\
If we denote
\ben
\Delta_k & = & \sum_{j=k+1}^L g_{kj}t_j,\nonumber\\
C_k & = & C - \sum_{i=k+1}^L g_{ii}\left(t_i + \sum_{j=i+1}^L g_{ij}t_j\right)^2,
\enn
the bounds for $s_k$ can be expressed as
\beq
LB_{k} \le t_{k} \le UB_{k},\label{sk1}
\enq
where
\ben
UB_{k} = \left\lfloor\; \sqrt{\frac{C_k}{g_{kk}}} - \Delta_k \;\right\rfloor, LB_{k} = \left\lceil\; -\sqrt{\frac{C_k}{g_{kk}}} - \Delta_k \;\right\rceil.\label{sk2}
\enn
Note that for given radius $\sqrt{C}$ and matrix ${\bf U}$, the bounds for $t_k$ only depends on the previous evaluated $t_{k+1}, t_{k+2}, \cdots, t_L$.

Finally, we evaluate the element $t_1$ of the searching vector ${\bf t}$. Referring to (\ref{arg_C}) and let $k=1$, we will have
\beq
\sum_{i=1}^{L} g_{ii} \left(t_i + \sum_{j=i+1}^L g_{ij}t_j\right)^2 \le C,
\enq
which leads to\\
\makebox
{\scalebox{0.8}{%
\parbox{0.5\textwidth}{
\ben
& \left\lceil\; -\sqrt{\frac{1}{g_{11}} \left(C - \sum_{i=2}^L g_{ii}\left(t_i + \sum_{j=i+1}^L g_{ij}t_j\right)^2\right)} - \sum_{j=2}^L g_{1j} t_j \;\right\rceil\nonumber\\
& \le t_1 \le \left\lfloor\; \sqrt{\frac{1}{g_{11}} \left(C - \sum_{i=2}^L g_{ii}\left(t_i + \sum_{j=i+1}^L g_{ij}t_j\right)^2\right)} - \sum_{j=2}^L g_{1j} t_j \;\right\rfloor.
\enn}}}\\
If we denote
\ben
\Delta_1 & = & \sum_{j=2}^L g_{1j}t_j,\nonumber\\
C_1 & = & C - \sum_{i=2}^L g_{ii}\left(t_i + \sum_{j=i+1}^L g_{ij}t_j\right)^2,
\enn
the bounds for $t_1$ can be expressed as
\beq
LB_{1} \le t_{1} \le UB_{1},\label{s11}
\enq
where
\ben
UB_{1} = \left\lfloor\; \sqrt{\frac{C_1}{g_{11}}} - \Delta_1 \;\right\rfloor,\quad LB_{1} = \left\lceil\; -\sqrt{\frac{C_1}{g_{11}}} - \Delta_1 \;\right\rceil.\label{s12}
\enn

In practice, $C_L$, $C_{L-1}$, $\cdots$, $C_1$ can be updated recursively by the following equations
\ben
\Delta_k & = & \sum_{j=k+1}^L g_{kj}t_j,\\
C_k & = & C - \sum_{i=k+1}^L g_{ii}\left(t_i + \sum_{j=i+1}^L g_{ij}t_j\right)^2\nonumber\\
& = & C_{k+1} - g_{k+1,k+1}\left(\Delta_{k+1} + t_{k+1}\right)^2,
\enn
for $k = L-1, L-2, \cdots, 1$ and $\Delta_L = 0$, $C_L = C$.

The entries $t_L, t_{L-1}, \cdots, t_1$ are chosen as follows: for a chosen candidate of $t_L$ satisfying the bounds (\ref{sL1})-(\ref{sL2}), we can choose a candidate for $t_{L-1}$ satisfying the bounds (\ref{sL-11})-(\ref{sL-12}). If a candidate value for $t_{L-1}$ does not exist, we go back to (\ref{sL1})-(\ref{sL2}) and choose other candidate value $t_L$. Then search for $t_{L-1}$ that meets the bounds (\ref{sL-11})-(\ref{sL-12}) for the given $t_L$. If $t_L$ and $t_{L-1}$ are chosen as candidates, we follow the same procedure to choose $t_{L-2}$, and so on. When a set of $t_L, t_{L-1}, \cdots, t_1$ is chosen and satisfies all corresponding bounds requirements, one candidate vector ${\bf t}=[t_1, t_2, \cdots, t_L]^T$ is obtained. We record all the candidate vectors satisfying their bounds requirements, such that all vectors meet  ${\bf t}^T {\bf G}_m {\bf t} \le C$ will be in $\Omega_m$.

Regarding the setting of positive constant $C$, we will set it based on the binary vector obtained by applying the direct sign operator of the real minimum-eigenvalue eigenvector of $\mathbf G_m$, denoted as $\mathbf t_{quant}$, such that
\beq
C = {\bf t}_{quant}^T {\bf G}_m\;{\bf t}_{quant}.\label{C}
\enq
By setting the searching sphere radius this way, it is big enough to have at least one searching vector $\mathbf t_{quant}$ falls inside, while in the meantime small enough to have not too many searching vectors within.

Note that this searching procedure will return {\em all} candidates that satisfy ${\bf t}^T {\bf G}_m {\bf t} \le C$. There is at least one candidate vector ${\bf t}_{quant}$ such that its entries satisfy all the bounds requirements. On the other hand, the maximum likelihood (ML) exhaustive search among all $\mathbf t\in\mathbb Z^L$, with optimal result ${\bf t}_{ML}$ that returns the minimum metric ${\bf t}^T {\bf G}_m {\bf t}$, or equivalently maximum the computation rate for one relay, will also fall inside the search bounds, since
\beq
{\bf t}_{ML}^T {\bf G}_m {\bf t}_{ML} \le {\bf t}_{quant}^T {\bf G}_m\;{\bf t}_{quant} = C.
\enq
Hence, we are guaranteed to include the local optimal network coding coefficient vector, which maximizes the computation rate for one relay $m$, in $\Omega_m^{T_{max}}$.

\begin{figure*}[!t]
\normalsize
\centerline{\psfig{file=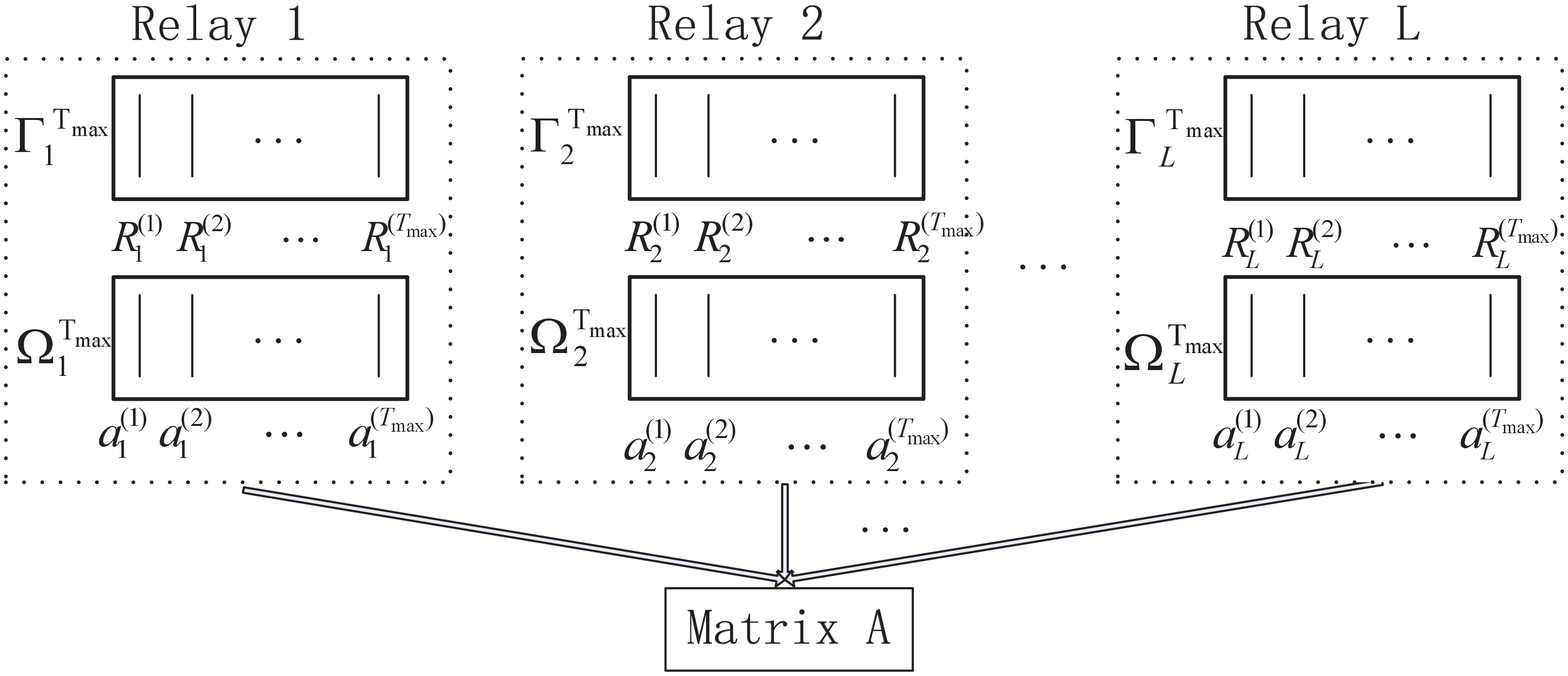,width=5in}}
\centering
\caption{Candidate sets and rate tables for all relays}
\hrulefill
\end{figure*}

We summarize our proposed algorithm for the searching candidate set $\Omega_m^{T_{max}}$ for relay $m$ based on Fincke-Pohst method as follows.

\vspace{-0.2cm}\hspace{-\parindent}\rule{\linewidth}{1pt}\vspace{-0.0cm}
\vspace{-0.0cm}{\bf Algorithm 1} FP Based Candidate Set Searching Algorithm\\
\rule{\linewidth}{.5pt}
{\em Input}: Matrix ${\bf G}_m$, $T_{max}=|\Omega_m^{T_{max}}|$.\\
{\em Output}: The candidate vector set $\Omega_m^{T_{max}}$ and corresponding computation rate set $\Gamma_m^{T_{max}}$.\\
\hspace{-\parindent}{\underline{Step 1}}: Calculate the binary quantized vector obtained by applying the direct sign operator of the real minimum-eigenvalue eigenvector of $\mathbf G_m$, denoted as $\mathbf t_{quant}$, and set $C$ as
\beq
C = {\bf t}_{quant}^T {\bf G}_m\;{\bf t}_{quant}.\label{Alg1_C}
\enq

\hspace{-\parindent}{\underline{Step 2}}: Operate Cholesky's factorization of matrix ${\bf G}_m$, ${\bf G}_m = {\bf U}^T {\bf U}$, where ${\bf U}$ is an upper triangular matrix. Let $u_{ij}$, $i,j=1, 2, \cdots,L$ denote the entries of matrix ${\bf U}$. Set
\beq
g_{ii}=u_{ii}^2, \quad\quad g_{ij}=u_{ij}/u_{ii},\nonumber
\enq
for $i = 1, 2, \cdots,L$, $j = i+1, \cdots, L$.

\hspace{-\parindent}{\underline{Step 3}}: Search set $\Omega_m = \left\{\mathbf t: \;\mathbf t^T \mathbf G_m \mathbf t\le C, \;\mathbf t\neq \mathbf 0,\;\mathbf t\in \mathbb Z^L\right\}$ according to the following Fincke-Pohst procedure.
\bei
\item[{\em (i)}] Start from $\Delta_L = 0$, $C_L = C$, $k = L$ and $\Omega_m=\emptyset$.
\item[{\em (ii)}] Set the upper bound $UB_{k}$ and the lower bound $LB_{k}$ as follows
\beq
UB_{k} = \left\lfloor\; \sqrt{\frac{C_k}{g_{kk}}} - \Delta_k \;\right\rfloor, LB_{k} = \left\lceil\; -\sqrt{\frac{C_k}{g_{kk}}} - \Delta_k \;\right\rceil,\nonumber
\enq
and $t_k = LB_k -1$.
\item[{\em (iii)}] Set $t_k = t_k + 1$. For $t_k \le UB_k$, go to (v); else go to (iv).
\item[{\em (iv)}] If $k=L$, terminate and output $\Omega_m$; else set $k = k + 1$ and go to (iii).
\item[{\em (v)}] For $k=1$, go to (vi); else set $k = k - 1$, and
\ben
\Delta_k & = & \sum_{j=k+1}^L g_{kj}t_j,\nonumber\\
C_k & = & C_{k+1} - g_{k+1,k+1}\left(\Delta_{k+1} + t_{k+1}\right)^2,\nonumber
\enn
then go to (ii).
\item[{\em (vi)}] If $\mathbf t = \mathbf 0$ terminate, else we get a candidate vector ${\bf t}\neq \mathbf 0$ that satisfies all the bounds requirements and put it inside $\Omega_m$, i.e. $\Omega_m = \{\Omega_m,\mathbf t\}$. Go to (iii).
\eni

\hspace{-\parindent}{\underline{Step 4}}: If $|\Omega_m| < T_{max}$, set $C = 2C$ and repeat \underline{Step 3}.

\hspace{-\parindent}{\underline{Step 5}}: Sort all the vectors $\mathbf t_1$, $\mathbf t_2$, $\cdots$, $\mathbf t_{|\Omega_m|}$ in $\Omega_m$ in descending order corresponding to the computation rate value $\mathscr R_m$ in (\ref{Rm}), such that
\beq
\mathscr R_m(\mathbf t_1) \ge \mathscr R_m(\mathbf t_2) \ge \cdots \ge \mathscr R_m(\mathbf t_{|\Omega_m|}).
\enq
Pick the first $T_{max}$ vectors of $\Omega_m$ to form the set $\Omega_m^{T_{max}}$ and construct the corresponding computation rate $\Gamma_m^{T_{max}}$ as
\ben
\left\{\begin{array}{lll}\Omega_m^{T_{max}} & = & \{\mathbf t_1,\mathbf t_2, \cdots, \mathbf t_{T_{max}}\},\\
\Gamma_m^{T_{max}} & = & \{\mathscr R_m(\mathbf t_1), \mathscr R_m(\mathbf t_2), \cdots, \mathscr R_m(\mathbf t_{T_{max}})\}.\end{array}\right.
\enn
\rule{\linewidth}{.5pt}

\subsection{Constructing Network Coding Matrix $\mathbf A$}

According to our proposed FP Based Candidate Set $\Omega_m^{T_{max}}$ Searching Algorithm 1, for relay $m$, we get the candidate set $\Omega_m^{T_{max}}$ for integer network coding coefficient vector $\mathbf a_m$. The set $\Omega_m^{T_{max}}$ consists $T_{max}$ candidates vectors $\Omega_m^{T_{max}} = \{\mathbf a_m^{(1)},\mathbf a_m^{(2)},\cdots,\mathbf a_m^{(T_{max})}\}$, in which $\mathbf a_m^{(1)}$, $\mathbf a_m^{(2)}$, $\cdots$, $\mathbf a_m^{(T_{max})}$ have been sorted such that $\mathscr R_m(\mathbf a_m^{(1)}) \ge \mathscr R_m(\mathbf a_m^{(2)}) \ge \cdots \ge \mathscr R_m(\mathbf a_m^{(T_{max})})$. Denote $\mathscr R_m^{(i)}=\mathscr R_m(\mathbf a_m^{(i)}), i=1,2,\cdots,T_{max}$. Then for each relay we can have two length-$T_{max}$ tables as shown in Fig. 4,
\ben
\textit{Table 1:} & &\Gamma_m^{T_{max}} = \{\mathscr R_m^{(1)},\mathscr R_m^{(2)},\cdots,\mathscr R_m^{(T_{max})}\},\\
\textit{Table 2:} & &\Omega_m^{T_{max}} = \{\mathbf a_m^{(1)},\mathbf a_m^{(2)},\cdots,\mathbf a_m^{(T_{max})}\}.
\enn
The second table consists the sorted candidate vector set $\Omega_m^{T_{max}}$, while the first one consists the corresponding computation rate set $\Gamma_m^{T_{max}}$ with elements $\mathscr R_m^{(1)} \ge \mathscr R_m^{(2)} \ge \cdots \ge \mathscr R_m^{(T_{max})}$.


After we get all the candidate vector sets $\Omega_1^{T_{max}}$, $\Omega_2^{T_{max}}$, $\cdots$, $\Omega_L^{T_{max}}$ and computation rate sets $\Gamma_1^{T_{max}}$, $\Gamma_2^{T_{max}}$, $\cdots$, $\Gamma_L^{T_{max}}$, we will try to pick up $\mathbf a_1\in \Omega_1^{T_{max}}$, $\mathbf a_2\in \Omega_2^{T_{max}}$, $\cdots$, $\mathbf a_L\in \Omega_L^{T_{max}}$, to construct the network coding system matrix $\mathbf A = [\mathbf a_1, \mathbf a_2, \cdots, \mathbf a_L]^T$ with full rank, while at the same time, the minimum corresponding rate $\mathscr R_1(\mathbf a_1)$, $\mathscr R_2(\mathbf a_2)$, $\cdots$, $\mathscr R_L(\mathbf a_L)$ is maximized.

Regarding this problem, first, we will sort the overall computation rate set for all relays $\{\Gamma_1^{T_{max}}, \Gamma_2^{T_{max}}, \cdots, \Gamma_L^{T_{max}}\}$ in a descending order into $\{\gamma_1, \gamma_2, \cdots, \gamma_{L\times T_{max}}\}$, such that $\gamma_1\ge \gamma_2\ge \cdots \ge \gamma_{L\times T_{max}}$. Then, starting from the largest possible achievable rate $\gamma_{\textit{index}}$ with $\textit{index}=L$ (the first $L-1$ rates are obviously not achievable), we will check one by one whether the rate $\gamma_{\textit{index}}$ is achievable, which means we can find $L$ vectors $\mathbf a_1\in \Omega_1^{T_{max}}$, $\mathbf a_2\in \Omega_2^{T_{max}}$, $\cdots$, $\mathbf a_L\in \Omega_L^{T_{max}}$, such that the following two constraints are satisfied:
\bei
\item[(i)] The system network coding coefficient matrix $\mathbf A$ is of full rank;
\item[(ii)] $\mathscr R_1(\mathbf a_1)$, $\mathscr R_2(\mathbf a_2)$, $\cdots$, $\mathscr R_L(\mathbf a_L)$ all greater or equal to $\gamma_{\textit{index}}$.
\eni
If not, we move to the next largest possible achievable rate $\gamma_{\textit{index}+1}$ and check in the same way, until the first achievable rate is found.

%

\begin{figure*}[!t]
\normalsize
\centerline{\psfig{file=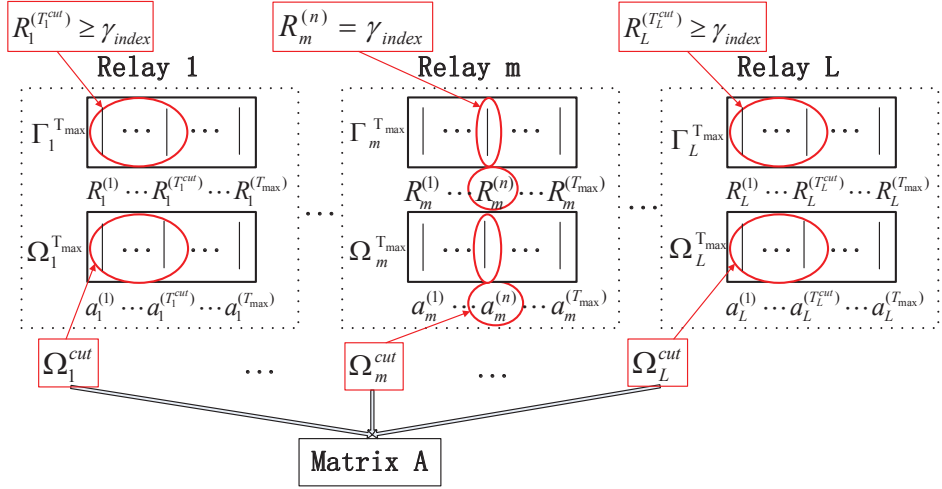,width=5in}}
\centering
\caption{Constructing network coding system matrix $\mathbf A$}
\hrulefill
\end{figure*}

When we are checking one possible achievable rate $\gamma_{\textit{index}}$, we will reduce/cut the network coding candidate set $\Omega_m^{T_{max}}$ into $\Omega_m^{cut}$ such that  any $\mathbf a_m\in\Omega_m^{cut}$ will satisfy that $\mathscr R_m(\mathbf a_m)$ greater or equal to $\gamma_{\textit{index}}$. In other words, the sets of $\Omega_1^{cut}$, $\Omega_2^{cut}$, $\cdots$, $\Omega_L^{cut}$ are constructed such that the constraint {\em (ii)} will definitely be satisfied if $\mathbf a_1\in\Omega_1^{cut}$, $\mathbf a_2\in\Omega_2^{cut}$, $\cdots$, $\mathbf a_L\in\Omega_L^{cut}$.

Suppose $\gamma_{\textit{index}}=\mathscr R_m^{(n)}\in \Gamma_m^{T_{max}}$, i.e. $\gamma_{\textit{index}}$ is taken from Table 1 of relay $m$ with table index $n$, then the network coding vector $\mathbf a_m^{(n)}$ is taken from Table 2 with the same index $n$, i.e. $\mathbf a_m^{(n)}\in \Omega_m^{max}$ is fixed for that relay and $\Omega_m^{cut}=\{\mathbf a_m^{(n)}\}$. For other relays $i\ne m$, the candidate set will reduce/cut to length $T_i^{cut}$ such that $\mathscr R_i^{(1)}$, $\mathscr R_i^{(2)}$, $\cdots$, $\mathscr R_i^{(T_i^{cut})}$ all greater or equal to $\gamma_{\textit{index}}$.

Denote $\Omega_i^{cut}=\{\mathbf a_i^{(1)},\mathbf a_i^{(2)},\cdots,\mathbf a_i^{(T_i^{cut})}\}$. We can start to check the constraint {\em (i)} of the system network coding matrix $\mathbf A$ constructed by any $\mathbf a_1\in\Omega_1^{cut}$, $\mathbf a_2\in\Omega_2^{cut}$, $\cdots$, $\mathbf a_L\in\Omega_L^{cut}$ . If there exists one constructed $\mathbf A$ with full rank, then this rate $\gamma_{\textit{index}}$ is achievable. The procedure is shown in Fig. 5.

We summarize this procedure to constructing the full rank network coding system matrix $\mathbf A$ with candidate sets $\Omega_1^{T_{max}}$, $\Omega_2^{T_{max}}$, $\cdots$, $\Omega_L^{T_{max}}$ and the corresponding computation rate sets $\Gamma_1^{T_{max}}$, $\Gamma_2^{T_{max}}$, $\cdots$, $\Gamma_L^{T_{max}}$ as follows.


\vspace{-0.2cm}\hspace{-\parindent}\rule{\linewidth}{1pt}\vspace{-0.0cm}
\vspace{-0.0cm}{\bf Algorithm 2}\\
Network Coding System Matrix Constructing Algorithm\\
\rule{\linewidth}{.5pt}
{\em Input}: Candidate vector sets $\Omega_1^{T_{max}}$, $\Omega_2^{T_{max}}$, $\cdots$, $\Omega_L^{T_{max}}$;\\
\hspace*{0.6cm} Computation rate sets $\Gamma_1^{T_{max}}$, $\Gamma_2^{T_{max}}$, $\cdots$, $\Gamma_L^{T_{max}}$.\\
{\em Output}: The network coding system matrix $\mathbf A$ constructed from $\mathbf a_1\in \Omega_1^{T_{max}}$, $\mathbf a_2\in \Omega_2^{T_{max}}$, $\cdots$, $\mathbf a_L\in \Omega_L^{T_{max}}$  with full rank that gives the maximum transmission rate $\mathscr R_D^{max}$.\\
\hspace{-\parindent}{\underline{Step 1}}: Sort the overall computation rate set for all relays $\{\Gamma_1^{T_{max}}, \Gamma_2^{T_{max}}, \cdots, \Gamma_L^{T_{max}}\}$ in a descending order into $\{\gamma_1, \gamma_2, \cdots, \gamma_{L\times T_{max}}\}$, such that $\gamma_1\ge \gamma_2\ge \cdots \ge \gamma_{L\times T_{max}}$. Initialize $\textit{index}=L$.

\hspace{-\parindent}{\underline{Step 2}}: Check whether the rate of $\gamma_{\textit{index}}$ is achievable by the following procedure. Suppose $\gamma_{\textit{index}}=\mathscr R_m^{(n)}\in \Gamma_m^{T_{max}}$. Then, for relay $i$, the reduced candidate set $\Omega_i^{cut}, i=1,2,\cdots,L$ will be constructed as follows.
\bei
\item[(i)] For relay $m$, set $\Omega_m^{cut}=\{\mathbf a_m^{(n)}\}$.
\item[(ii)] For relay $i\neq m$, compare the value of $\gamma_{\textit{index}}$ and the sorted descending set $\Gamma_i^{T_{max}}=\{\mathscr R_i^{(1)}$, $\mathscr R_i^{(2)}$, $\cdots$, $\mathscr R_i^{(T_{max})}\}$. Find all $\{\mathscr R_i^{(1)}$, $\mathscr R_i^{(2)}$, $\cdots$, $\mathscr R_i^{(T_i^{cut})}\}$ greater or equal to $\gamma_{\textit{index}}$. Set $\Omega_i^{cut}=\{\mathbf a_i^{(1)}$, $\mathbf a_i^{(2)}$, $\cdots$, $\mathbf a_i^{(T_i^{cut})}\}$.
\eni

\hspace{-\parindent}{\underline{Step 3}}: Check every $\mathbf a_1\in \Omega_1^{cut}$, $\mathbf a_2\in \Omega_2^{cut}$, $\cdots$, $\mathbf a_L\in \Omega_L^{cut}$, until we find one network coding system matrix $\mathbf A = [\mathbf a_1, \mathbf a_2, \cdots, \mathbf a_L]^T$ has full rank, i.e. $|\mathbf A|\neq 0$. If so, terminate and output the network coding system matrix $\mathbf A$ and the maximum transmission rate $\mathscr R_D^{max}=\gamma_{\textit{index}}$.

\hspace{-\parindent}{\underline{Step 4}}: If for any $\mathbf a_1\in \Omega_1^{cut}$, $\mathbf a_2\in \Omega_2^{cut}$, $\cdots$, $\mathbf a_L\in \Omega_L^{cut}$, we cannot construct a full rank network coding system matrix $\mathbf A$, then set $\textit{index}=\textit{index}+1$, go to Step 2. \\
\rule{\linewidth}{.5pt}

One possible implementation of the whole system will let relays calculate the candidate sets and corresponding computation rate sets, construct the optimal network coding system matrix $\mathbf A$, then transmit the $L\times L$ integers matrix $\mathbf A$ to the destination. Another possible implementation is to allow the destination work as processing center, that does all calculations, including candidate sets, corresponding computation rate sets, and the optimal network coding system matrix $\mathbf A$ construction. The destination will then feedback the optimal network coding vector $\mathbf a_m\in\mathbb Z^L$ to relay $m$ for $m=1,2,\cdots,L$. After system initialization, these optimal network coding vectors can be used for the system when the channels are stationary.

\section{Experimental Studies}

\subsection{A Transparent Realization}

In this subsection, we will give a detailed experimental example to show our proposed algorithms in a transparent way. For a three-source three-relay system with $L=3$, we set the power constraints $P=10dB$ and $T_{max}=5$. The channel coefficient vector $\mathbf h_m$ for each relay is generated as
\ben
\mathbf h_1 & = & [0.9730,0.4674,0.5103]^T,\nonumber\\
\mathbf h_2 & = & [-1.7291,0.7166,-0.5856]^T,\nonumber\\
\mathbf h_3 & = & [-0.3912,1.4407,-0.8115]^T.\nonumber
\enn

After calculating $\mathbf G_m$, $m=1,2,3$ and running our proposed FP based candidate set searching algorithm for each relay, we will get the network coding candidate vector sets $\Omega_1^{T_{max}}$, $\Omega_2^{T_{max}}$, $\Omega_3^{T_{max}}$ and corresponding computation rate sets $\Gamma_1^{T_{max}}$, $\Gamma_2^{T_{max}}$, $\Gamma_3^{T_{max}}$ as follows
\ben
\Omega_1^{T_{max}} & = & \left[\begin{array}{ccccc}
1&2&1&1&1\\
0&1&1&0&1\\
0&1&1&1&0\end{array}\right],\nonumber\\
\Gamma_1^{T_{max}} & = & \left[0.4846,\;\; 0.4620,\;\; 0.3408,\;\; 0.2918,\;\; 0.2231\right];\nonumber
\enn
\ben
\Omega_2^{T_{max}} & = & \left[\begin{array}{ccccc}
1&2&3&-1&-2\\
0&-1&-1&1&1\\
0&1&1&0&0\end{array}\right],\nonumber\\
\Gamma_2^{T_{max}} & = & \left[0.7087,\;\; 0.6785,\;\; 0.5572,\;\; 0.3625,\;\; 0.2694\right];\nonumber
\enn
\ben
\Omega_3^{T_{max}} & = & \left[\begin{array}{ccccc}
0&0&1&0&1\\
-1&1&-2&-2&-3\\
1&0&1&1&2\end{array}\right],\nonumber\\
\Gamma_3^{T_{max}} & = & \left[0.5987,\;\; 0.5935,\;\; 0.4384,\;\; 0.4165,\;\; 0.2902\right].\nonumber
\enn
We can see that the computation rate set $\Gamma_m^{T_{max}}$, $m=1,2,3$ has elements sorted in descending order where the first element is the maximum computation rate for relay $m$. The $n$-th column in $\Omega_m^{T_{max}}$ is a candidate network coding vector $\mathbf a_m^{(n)}$ for relay $m$, while the corresponding computation rate is the $n$-th element in $\Gamma_m^{T_{max}}$. Note that if we optimize the network coding coefficients separately, which means each relay will use network coding vector that maximizes its own computation rate, $\mathbf a_m$ is taken from the first column of $\Omega_m^{T_{max}}$, $m=1,2,3$ and the constructed network coding system matrix
\beq
\mathbf A_{\textit{local}} = \left[\begin{array}{ccc}
1&1&0\\
0&0&-1\\
0&0&1\end{array}\right]^T\nonumber
\enq
is obviously not of full rank. In this case, the destination actually cannot decode all the messages efficiently.

Then we go forward to run our proposed network coding system matrix constructing algorithm. We sort the computation rates for all relays in a descending order,
\beq
\{\underbrace{0.7087}_{\gamma_1},\underbrace{0.6785}_{\gamma_2},\underbrace{0.5987}_{\gamma_3},\underbrace{0.5935}_{\gamma_4},\underbrace{0.5572}_{\gamma_5},\underbrace{0.4846}_{\gamma_6},\cdots\}.\nonumber
\enq
and start to check the rate from the third maximum value, $\gamma_3 = 0.5987$, then $\gamma_4=0.5935$, then $\gamma_5=0.5572$, $\cdots$, to see whether it is achievable. If so, terminate and output; if not, move to the next rate.

For example, when we are checking $\gamma_4=0.5935=\mathscr R_3^{(2)}$, which is taken from the second element of $\Gamma_3^{T_{max}}$, the reduced candidate sets $\Omega_1^{cut}$, $\Omega_2^{cut}$, $\Omega_3^{cut}$ with all corresponding rates greater or equal to $\gamma_4=0.5935$ can be constructed as
\ben
\Omega_1^{cut} = \emptyset,\quad\quad \Omega_2^{cut} = \left[\begin{array}{cc}
1&2\\
0&-1\\
0&1\end{array}\right],\quad\quad \Omega_3^{cut} = \left[\begin{array}{c}0\\1\\0\end{array}\right].\nonumber
\enn
We can easily see that no full rank network coding system matrix $\mathbf A$ can be constructed with $\mathbf a_1\in \Omega_1^{cut}$, $\mathbf a_2\in \Omega_2^{cut}$, $\mathbf a_3\in \Omega_3^{cut}$. Hence the rate of $\gamma_4=0.5935$ is not achievable. We will move to $\gamma_5=0.5572$ and check in the same way.

After running our proposed Network Coding System Matrix $\mathbf A$ Constructing Algorithm 2, the network coding system matrix $\mathbf A = \left[\mathbf a_1, \mathbf a_2, \mathbf a_3\right]^T$ is finally constructed as
\beq
\mathbf A_{\textit{proposed}} = \left[\begin{array}{ccc}
1&2&0\\
0&-1&1\\
0&1&0\end{array}\right]^T\nonumber
\enq
and the maximum transmission rate $\mathscr R_D^{max}=0.4846$.

\subsection{Simulation Results}

We present numerical results to evaluate the performance of our proposed algorithms. First, we show that if network coding integer coefficient vector is optimized separately/locally at each relay, the probability that the network coding system matrix $\mathbf A$ is not of full rank, i.e. $|\mathbf A|=0$, in which case the destination actually cannot decode the original messages efficiently. With the average of $10000$ randomly generated channel realizations, it can be observed from Fig. 6 the severity of this issue. For example, when $L=3$ and $P=1$dB-$8$dB, the probability of rank failure with local optimized network coding vectors is always beyond $0.4$. This further assures the importance and necessity of our proposed algorithms.

\begin{figure}[hbt]
\centerline{\psfig{file=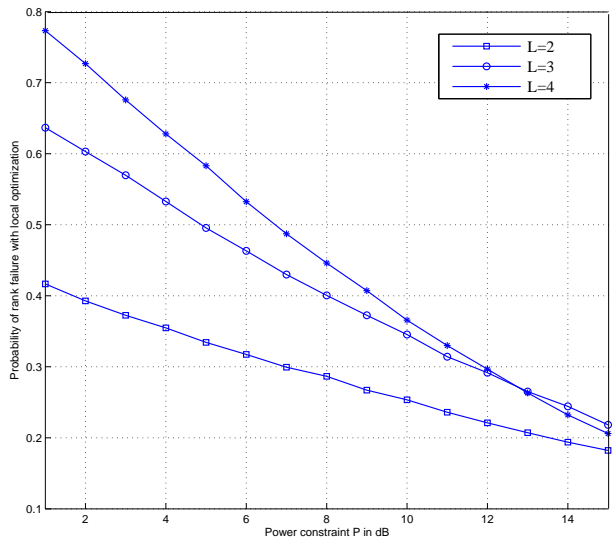,width=3.8in}}
\centering
\caption{Probability of rank failure with local optimization for $L=2,3,4$}
\label{figure-percentage}
\end{figure}

In Fig. 7, we compare the overall transmission rate $\mathscr R_D$ at destination, with the average of $10000$ randomly generated channel realizations, of several different strategies in multi-source multi-relay channels with $L=3$ and $T_{max}=5$. (i) The ``{\em DF with interference as noise}'' is a strategy in which relay $m$ is trying to decode one message from source $m$ and treat other messages as noise. In this special case, the system matrix $\mathbf A=\mathbf I_L$. (ii) The ``{\em CPF NC with Round-H}'' is a strategy that each relay decodes a linear integer combination of transmitted messages, while the network coding coefficients are set by a simplified method, i.e. rounding the channel coefficients directly to the nearest integers. (iii) The ``{\em CPF NC with local optimization}'' is a strategy that each relay also decodes a linear integer combination of transmitted messages, while the network coding coefficients are optimized locally/separately. Due to the rank failure issue of network coding system matrix, in which case the destination cannot decode all messages, the rate is decreased. Finally, (iv) the ``{\em CPF NC with proposed algorithms}'' is the strategy that each relay decodes a linear integer combination of transmitted messages with our proposed FP based candidate set searching algorithm and network coding system matrix constructing algorithm.

\begin{figure}[hbt]
\centerline{\psfig{file=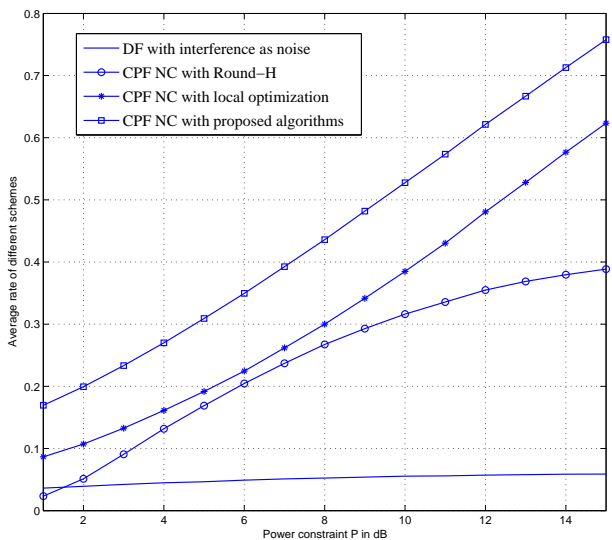,width=3.8in}}
\centering
\caption{Rate comparisons of different schemes for $L=3$}
\label{figure-simulation1}
\end{figure}

As shown in Fig. 7, the performance differences are significant. ``{\em DF with interference as noise}'' gives very poor result. Furthermore, increasing power constraint has not much effect on this strategy since as the power increases for the interested message, the corresponding interference power is also raised. The ``{\em CPF NC with Round-H}'' strategy works a little better since it somehow takes advantage of network coding to improve the rate, but the coefficients are chosen in a simplified way and not optimal. The ``{\em CPF NC with proposed algorithms}'' strategy, in which case the network coding coefficients are optimized systematically, performs superior to all other strategies and has about $3$dB gain compared with the ``{\em CPF NC with local optimization}''.

\begin{figure}[hbt]
\centerline{\psfig{file=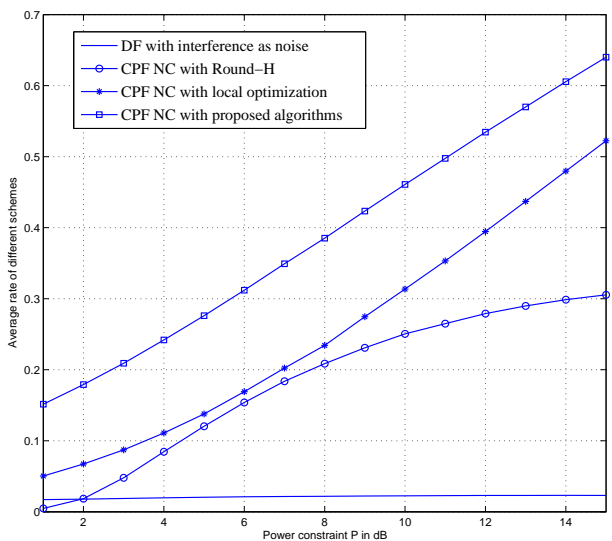,width=3.8in}}
\centering
\caption{Rate comparisons of different schemes for $L=4$}
\label{figure-simulation2}
\end{figure}

We repeat our experiment with multi-source multi-relay channels of $L=4$ and present the average rate comparisons of different schemes with respect to the power constraint. Similar results are shown as in Fig. 8. ``{\em CPF NC with proposed algorithms}'' strategy still gives the best performance and further demonstrates the effectiveness of our proposed algorithms.

\section{Conclusion}
In this work, we consider the problem of integer network coding coefficients design in a system level over a compute-and-forward multi-source multi-relay system. Instead of optimizing network coding vector of each relay separately, we propose the Fincke-Pohst based candidate set searching algorithm, to provide a network coding vector candidate set for each relay with corresponding computation rate in descending order. Then, with our proposed network coding system matrix constructing algorithm, we choose network coding vectors from candidate sets to construct network coding system matrix with full rank, while in the meantime the transmission rate of the overall system is maximized. Numerical results give the performance comparisons of our proposed compute-and-forward network coding algorithms and other strategies.

\end{document}